\documentclass[doublecol]{epl2}

\usepackage{amsmath, amssymb, color, stmaryrd, wasysym, esint}
\usepackage{scalerel}
\usepackage{cuted} 

\makeatletter
\newsavebox{\@brx}
\newcommand{\llangle}[1][]{\savebox{\@brx}{\(\m@th{#1\langle}\)}%
  \mathopen{\copy\@brx\kern-0.5\wd\@brx\usebox{\@brx}}}
\newcommand{\rrangle}[1][]{\savebox{\@brx}{\(\m@th{#1\rangle}\)}%
  \mathclose{\copy\@brx\kern-0.5\wd\@brx\usebox{\@brx}}}
\makeatother

\makeatletter
\newsavebox{\@brxx}
\newcommand{\lllangle}[1][]{\savebox{\@brxx}{\(\m@th{#1\langle}\)}%
  \mathopen{\copy\@brxx\kern-0.5\wd\@brxx\usebox{\@brxx}\kern-0.5\wd\@brxx\usebox{\@brxx}}}
\newcommand{\rrrangle}[1][]{\savebox{\@brxx}{\(\m@th{#1\rangle}\)}%
  \mathclose{\copy\@brxx\kern-0.5\wd\@brxx\usebox{\@brxx}\kern-0.5\wd\@brxx\usebox{\@brxx}}}
\makeatother

\usepackage{pdfpages}
\usepackage{graphicx}
\usepackage{dcolumn}
\usepackage{bm}
\usepackage{xcolor}
\usepackage{xspace}
\usepackage[makeroom]{cancel}
\usepackage{float}
\usepackage{siunitx}
\usepackage{mathtools}
\usepackage{algorithm}
\usepackage{algorithmic}

\definecolor{linkcolor}{rgb}{0,0,0.6} 
\usepackage[colorlinks=true,
    pdfstartview = FitV,
    linkcolor    = linkcolor,
    citecolor    = linkcolor,
    urlcolor     = linkcolor,
    hyperindex   = true,
    hyperfigures = false]{hyperref}

\title{Geometric decomposition of the $d$-dimensional hard-sphere partition function}
\shorttitle{Geometric decomposition of the hard-sphere partition function}

\author{Luke K. Davis\inst{1,2}\thanks{E-mail: \email{luke.davis@ed.ac.uk}}}
\shortauthor{L. K. Davis}

\institute{                    
  \inst{1} School of Mathematics and Maxwell Institute for Mathematical Sciences, University of Edinburgh, EH9 3FD, Scotland\\
  \inst{2} Higgs Centre for Theoretical Physics, University of Edinburgh, EH9 3FD, Scotland
}

\renewcommand{\doi}[1]{}

\abstract{
We introduce a geometric decomposition of the hard-sphere partition function. Using a close-packing-inspired geometric bound on the available insertion volume, made rigorous when a corresponding local density certificate is available, we establish a reference upper bound $Q^\ast$ on the configurational integral. Factoring this upper bound out of the statistical geometric partition function found by Speedy yields a new form for the $d$-dimensional partition function, $Q(N,V,T)=Q^\ast \exp(-N \mathcal{J})$, where $\mathcal{J}$ depends strictly on the boundary-to-volume ratio of the voids and the close-packing density. Overall, this work provides an exact reformulation of hard-sphere thermodynamics, thus deepening our statistical geometric understanding of hard spheres.
}

\begin{document}

\maketitle

\section{Introduction}
A grand challenge in statistical mechanics is to understand fully the equilibrium hard-sphere system, where no two spheres come closer than the sum of their radii, across the full range of densities and for all spatial dimensions \cite{maxwell1890,boltzmann1896,Tonks1936,Rice1942,Reiss1959,Alder1959,wertheim1963,thiele1963,Carnahan1969,Frenkel1984,Frisch1985,Lowen,Parisi2006,Parisi2010,Royall2024}. Despite continual effort, the thermodynamics of the hard-sphere system has only been fully (exactly) solved in one-dimension \cite{Tonks1936} and in infinite dimensions \cite{Frisch1985,Parisi2006,Kurchan2012}, with all the dimensions in between remaining elusive.

The athermal nature of the hard-sphere model means all of its properties are geometric. The properties of the fluid are, moreover, averaged quantities: one ought to account for all the possible arrangements of the spheres. The marrying, and exploration, of these essential facts have been the concern of statistical geometry \cite{Beresford1969,Speedy1977,Speedy1980,Reiss1986,Speedy1988,Reiss1989,Reiss1992,Speedy1991,Sastry1997,Sastry1998,Bowles2000}. From statistical geometric considerations a few exact relationships between thermodynamic state functions and geometry have been derived, mainly by R. J. Speedy \cite{Speedy1980,Speedy1991}. The main geometric quantities are the insertion volume ($V_I$), the total volume that is available to insert another particle without overlaps, the boundary of this volume ($S_I$), and their free-volume analogues: the free-volume ($v_f$) and surface area ($s_f$) that is accessible to a particle performing continuous motion from its current position \cite{hoover1972exact,hoover1979exact,Sastry1998}. These quantities are all underpinned by the union of arrangements of congruent balls in $d$ dimensions \cite{rogers1964packing,Conway1999,csikos1998,bezdek2025}, which still has no closed-form solution though exact, and otherwise accurate and efficient, algorithms exist to compute them \cite{edelsbrunner1995,Sastry1997,Davis2026IS}.

Here, we present a reformulation of the exact partition function of the $d$-dimensional hard-sphere system based upon statistical-geometric bounds. The purpose of the present paper is to show, and interpret, the following:\\

\textbf{Theorem 1.--} The exact $d$-dimensional canonical partition function $Q_N \equiv Q(N,V,T)$ for $N$ hard spheres of diameter $\sigma$ in a volume $V$ at temperature $T$ admits the form
\begin{align}
Q_N &= Q^\ast_{N} \exp(-N \mathcal{J}) , \\
    Q^\ast_{N} &= \frac{1}{\Lambda^{Nd}} \left( \frac{v_d}{\Delta(d)} \right)^N \frac{1}{N!} \frac{\Gamma\left( \frac{V \Delta(d)}{v_d} + 1 \right)}{\Gamma\left( \frac{V \Delta(d)}{v_d} - N + 1 \right)},
\end{align}
where $\Lambda(T)$ is the thermal de-Broglie wavelength, $v_d$ is the volume of the $d$-dimensional ball of diameter $\sigma$,  $\Delta(d) \equiv \Delta_d \ge \phi_{cp}(d)$ is a rigorously certified density scale (with $\phi_{cp}(d)$ the close-packing density of hard spheres in $d$ dimensions), and $\Gamma(\cdot)$ is the Gamma function. The finite-$N$ geometric penalty is exactly defined as $\mathcal{J}_{N} = -\frac{1}{N} \sum_{i=0}^{N-1} \ln [ \mathcal{P}_I(\phi_i) / (1-\phi_i/\Delta_d) ]$, with $\phi_i=i v_d/V$ and $\mathcal{P}_I(\phi_i)$ being the insertion probability. In the thermodynamic limit $\mathcal{J}_{N} \to \mathcal{J}(\phi)$, where
\begin{equation}
     \mathcal{J}(\phi) =  \frac{\sigma}{2d} \int_0^\phi \frac{\langle S_I(\phi') \rangle}{\langle V_I(\phi') \rangle} \frac{d\phi'}{\phi'}  -1 -\frac{\Delta_d-\phi}{\phi} \ln\left(1 - \frac{\phi}{\Delta_d} \right).
    \label{eq:J_theorem}
\end{equation}

We will later show that the sharp form of Theorem 1 involving $\Delta_d = \phi_{cp}(d)$ is strictly rigorous in any dimension where a corresponding (sharp) Blichfeldt-type local density certificate can be supplied. It will be conducive, for the physics, to revert back to expressing quantities through $\phi_{cp}$. The integrand $\langle S_I (\phi')\rangle/\langle V_I (\phi') \rangle$ may also be expressed in terms of $\langle s_f/v_f \rangle(\phi')$, \emph{i.e.,} the ensemble-averaged ratio of the local (cavity) free-volume area to the local free-volume \cite{Speedy1991,Sastry1998}.

A sketch of our formulation is as follows: we first find the extremal geometry of hard spheres, inspired by the Kneser-Poulsen contraction problem \cite{poulsen1954,kneser1955,csikos1998}, which asymptotically saturates the insertion-volume bound. We then, using a partition function relation from Speedy \cite{Speedy1980}, provide an upper bound on the hard-sphere partition function $Q^\ast$. The final form then results from factoring out this upper bound from the configurational integral. We stress that what we provide is an exact geometric factorization of the hard-sphere partition function, valid for all $N$ and $\phi$, with the hard-sphere system interpreted as a “fragmented void-space” correction to a close-packed, connected, droplet reference. The many-body content of the problem is thereby relocated, not solved, into $\mathcal{J}$, whose evaluation still requires the equilibrium input $\langle S_I \rangle/\langle V_I \rangle$.

\section{Statistical insertion geometry}
We consider systems of $N$ monodisperse hard spheres of diameter $\sigma$ residing in a $d$-dimensional flat torus (a hypercube with periodic boundary conditions) of volume $V=L^d$. The standard hard-sphere non-overlap constraint is:
\begin{equation}
    | \mathbf{r}_i - \mathbf{r}_j | \geq \sigma \qquad \forall \quad i \neq j.
    \label{eq:HSconstraint}
\end{equation}

We denote a $d$-dimensional ball parameterized by its diameter $\sigma$ as $B_\sigma(d)$. The physical volume of a single hard sphere is thus $v_d = \text{Vol}(B_\sigma(d))$ and is explicitly determined as
\begin{equation}
\text{Vol}(B_\sigma(d)) = \frac{\pi^{d/2} \sigma^d}{2^d\Gamma(1+d/2)},
\end{equation}
and the global packing fraction is defined as $\phi = N v_d / V$.

A direct implication of  \eqref{eq:HSconstraint} is the presence of an exclusion zone: the spherical space around each particle that is strictly inaccessible to the {center} of any to-be inserted sphere. Because the centers of two non-overlapping spheres of diameter $\sigma$ cannot be closer than a distance $\sigma$, this exclusion ball possesses a diameter of $2\sigma$. This exclusion ball, centered at $\mathbf{r}_i$, is denoted $B_{2\sigma}(\mathbf{r}_i; d)$ and has a volume of $2^d v_d$.

For a configuration $\omega$, obeying \eqref{eq:HSconstraint}, the space available to insert another sphere is precisely the set of points lying outside these exclusion balls:
\begin{align}
    V_I(\omega,N;d) &= V - \text{Vol}\left(\bigcup_{i=1}^N B_{2\sigma}(\mathbf{r}_i ;d)\right),  \label{eq:defV_I} \\
    &= \int_V d \mathbf{r}  \left[ 1 - \Theta\left(\sum_{i=1}^N \Theta(\sigma - |\mathbf{r} - \mathbf{r}_i|) \right) \right],
    \label{eq:defV_I2}
\end{align}
where $\Theta(\cdot)$ is the Heaviside step function. The probability of successful insertion of another sphere is simply $\mathcal{P}_I(\omega,N;d) = V_I(\omega,N) / V$ \cite{Widom1963}.

\begin{figure}[t!]
    \centering
    \includegraphics[width=1.0\linewidth]{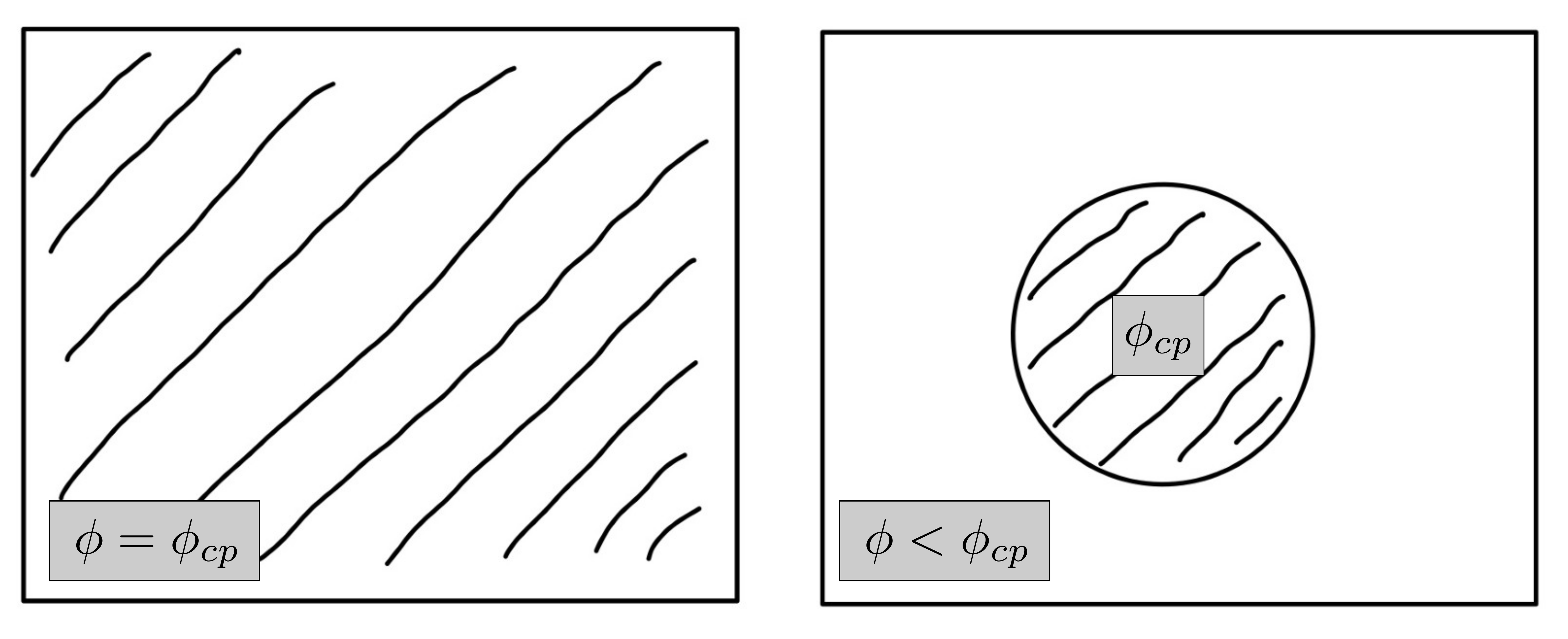}
    \caption{ Illustrations of a globally close-packed system (left) and a close-packed droplet reference (right), with a global packing fraction less than the close-packing fraction. In the thermodynamic limit this construction (right) asymptotically saturates the insertion-volume upper bound.}
    \label{fig:densecluster}
\end{figure}

The boundary of the insertion space is the outer envelope of the union of these exclusion balls. Using the $(d-1)$-dimensional Hausdorff measure, the surface area of the insertion space is:
\begin{equation}
    S_I(\omega, N; d) = \mathcal{H}^{d-1}\left( \partial \bigcup_{i=1}^N B_{2\sigma}(\mathbf{r}_i; d) \right).
\end{equation}

\textit{Upper bounds for $V_I$:-} We now ask, for a given $N$, $d$, and $V$, is there a configuration, say $\omega^\ast$, which (asymptotically) saturates the insertion volume? From \eqref{eq:defV_I}, maximising the insertion volume $V_I$ is equivalent to minimizing the volume of the union of the exclusion balls. Thus, finding $\max_\omega V_I$ is the same as performing $\min_\omega \int_V d \mathbf{r} \Theta(\theta(\mathbf{r}))$, where $\theta(\mathbf{r})=\sum_{i=1}^N \Theta(\sigma - \vert{}\mathbf{r} - \mathbf{r}_i\vert{})$ is a covering field. Since $\int_V d \mathbf{r} \theta(\mathbf{r}) = N 2^d v_d$ is a conserved field, the mechanism for asymptotically saturating $V_I$ is to minimize the spatial support of the $\theta(\mathbf{r})$ by increasing its local amplitude. Physically, this extremisation forces the exclusion balls to intersect as much as possible, within the constraint of \eqref{eq:HSconstraint}, so that the union is minimal.

The geometry of the special configuration $\omega^\ast$ would then be the tightest contraction of all the ball centers, \emph{i.e.,}
$\sum_{i \neq j} |\mathbf{r}_i - \mathbf{r}_j|(\omega^\ast) \leq  \sum_{i \neq j} |\mathbf{r}_i - \mathbf{r}_j|(\omega),$
which naturally would be a $d-$dimensional spherical cluster (see Fig. \ref{fig:densecluster}). Note, while this physical picture is intuitively connected to the Kneser-Poulsen contraction conjecture \cite{poulsen1954,kneser1955,Csikos2018,bezdek2025}, that for a contraction (of the ball centers) the union of balls does not increase, the resulting thermodynamic bound on the insertion probability can be established independently of it (see Appendix A in the Supplementary Information (SI)).

For the special configuration $\omega^\ast$, all $N$ particles form a single spherically-shaped cluster of volume $V^\ast = V \phi / \phi_{cp}(d)$, where $\phi_{cp}(d)$ is the close-packed density of spheres in that dimension. The radius of this  cluster, $R^\ast$, is obtained from the total physical volume of the spheres:
\begin{equation}
    R^\ast = \frac{\sigma}{2} \left( \frac{N}{\phi_{cp}(d)} \right)^{1/d}. \label{eq:ClusterRadius}
\end{equation}

In this tight cluster picture, the excluded-volume, and its complement: the insertion volume, form fully-connected regions. So, within the bulk of this dense cluster the exclusion balls overlap so as to cover the local space and leave no insertion holes. For this ``macroscopic'' spherical-droplet reference, the exclusion union is, up to surface-order corrections, a ball of radius $R^\ast+\sigma$, giving
\begin{equation}
 \text{Vol}\left(\bigcup_{i=1}^{N} B_{2\sigma}(\mathbf{r}_i ;d)\right) \geq V^\ast \left( 1 + \frac{\sigma}{R^\ast} \right)^d,
\end{equation}
the right-most factor accounts for a thin exclusion layer extending an additional radial distance $\sigma$ beyond the outermost layer of sphere centers. The total volume of this minimized union of exclusion balls is therefore the volume of a $d$-dimensional ball of radius $R^\ast + \sigma$. 

For the spherical close-packed droplet reference, the insertion volume is the remaining hyper-volume outside this exclusion halo $
    V_I(N) = V - V^\ast \left( 1 + \sigma/R^\ast \right)^d.$
Expanding this finite-size boundary correction yields $V_I(N) \simeq V - V^\ast - d V^\ast (\sigma/R^\ast)$. Because the cluster volume $V^\ast$ scales extensively with $N$, while the cluster radius scales as $R^\ast \propto N^{1/d}$, the volume of this exclusion layer scales sub-extensively as an area, $\mathcal{O}(N^{(d-1)/d})$. 

In the thermodynamic limit ($N, V \to \infty$ at fixed $\phi$), because $\sigma / R^\ast \propto N^{-1/d}$, the thin exclusion layer strictly vanishes resulting in:
\begin{align}
\mathcal{P}_I(\phi;d) &\leq \frac{V-V^\ast}{V} = 1 - \frac{\phi}{\phi_{cp}(d)}. \label{eq:PI_upper}
\end{align}
While the spherical cluster provides a physically motivated macroscopic reference state, establishing this maximum from contraction arguments is mathematically subtle. In Appendix A (SI), we show that Eq.~\eqref{eq:PI_upper} follows rigorously whenever a sharp Blichfeldt-type local density certificate \cite{rogers1964packing} exists. More generally, if a Blichfeldt-type local density certificate at density scale $\Delta_d$ is available, it gives the certified bound $\mathcal{P}_I \le 1 - \phi/\Delta_d$.

The sharp form involving $\phi_{cp}(d)$ is rigorous wherever such a certificate can be supplied; this is presently known explicitly only for $d=1,8,24$, via the linear-programming construction of Viazovska and others \cite{Viazovska2017,Cohn2017}. Although $\phi_{cp}(d)$ is also known rigorously for $d=2,3$ \cite{Toth1950,Hales2005}, these results follow from local cell-decomposition arguments rather than an explicit certificate, and whether the bound can be sharpened there remains, to our knowledge, open. In any case, our formalism remains fully rigorous using any certified density $\Delta_d \ge \phi_{cp}(d)$ (e.g. the Kabatiansky-Levenshtein bound below).

For asymptotically large dimensions ($d \to \infty$), the close-packing fraction is not exactly known. However, Kabatiansky and Levenshtein established an upper bound on the packing fraction of hard spheres \cite{Kabatiansky1978}, proving that $\phi_{cp}(d)$ decays exponentially as $d$ increases, bounded by $\phi_{cp}(d) \lesssim 2^{-d\times0.5990\ldots}$. If a local density certificate with the Kabatiansky-Levenshtein asymptotic density scale is available, the corresponding insertion bound would read
$\mathcal{P}_I(\tilde{\phi}; d) \le 1 - \tilde{\phi} \, 2^{-d\times0.4010\ldots}$,
where $\tilde{\phi} = 2^d \phi$ is the scaled packing fraction. This relation demonstrates that as $d\rightarrow \infty$, at a constant scaled density, the upper bound of the probability of insertion approaches 1; being certificate-based at a generic scale $\Delta_d$, this is consistent with the general, non-sharp, form of Theorem 1.

A simple lower bound for the insertion volume $V_I$ is obtained by identifying the configuration which maximizes the union of the exclusion balls. By Boole's inequality, or subadditivity of the Lebesgue measure, the volume of the union of any finite set of measurable regions is bounded from above by the sum of the individual volumes
\begin{equation}
    \text{Vol}\left(\bigcup_{i=1}^N B_{2\sigma}(\mathbf{r}_i ;d)\right) \leq \sum_{i=1}^N \text{Vol}(B_{2\sigma}(\mathbf{r}_i ;d)).
\end{equation}
\begin{figure}[t!]
    \centering
    \includegraphics[width=0.99\linewidth]{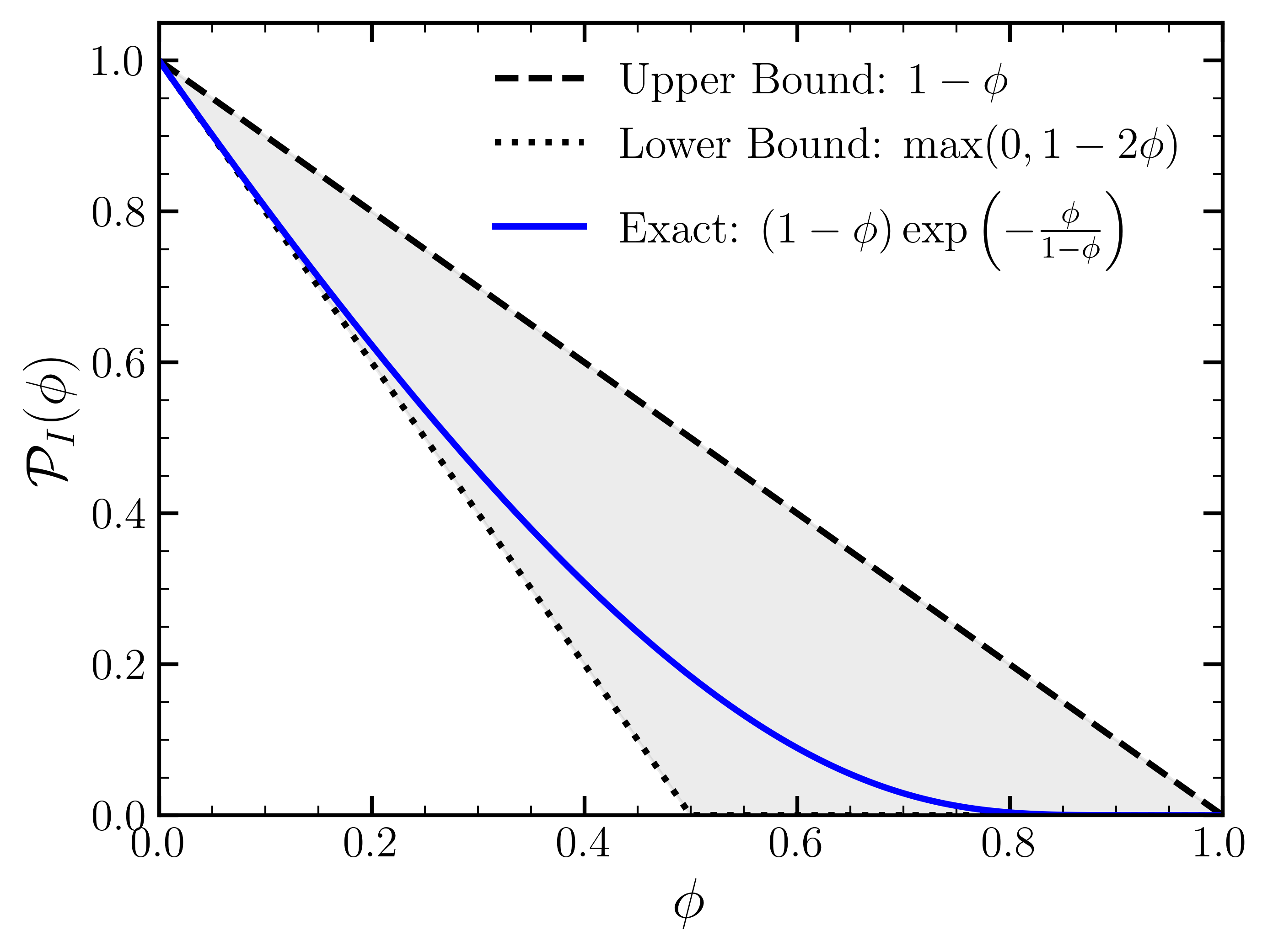}
    \caption{Comparing bounds on the insertion probability $\mathcal{P}_I = \langle V_I(\phi)\rangle/V$ with the exact solution for the $d=1$ hard-sphere system \cite{Tonks1936,Speedy1980,Speedy1991}.}
    \label{fig:PI-1d}
\end{figure}

Thus, the sought-for configuration $\tilde{\omega}$ is for balls to be well-separated, \emph{i.e,}
\begin{equation}
    \sum_{i \neq j} |\mathbf{r}_i - \mathbf{r}_j|(\tilde{\omega}) \geq  \sum_{i \neq j} |\mathbf{r}_i - \mathbf{r}_j|(\omega),
\end{equation}
and since the volume of a single exclusion ball is $2^d v_d$, the sum of all exclusion balls is at most $2^d \phi V$. Then, the insertion probability is bounded from below as
\begin{equation}
    \mathcal{P}_I(\phi;d) \geq \max(0, 1 - 2^d \phi),
    \label{eq:PI_lowerBound}
\end{equation}
which is tight at low densities where the configuration $\tilde{\omega}$ is realizable. As the packing fraction increases away from the dilute regimes the exclusion balls tend to overlap. To maintain the absolute minimum of insertion probability at these packing fractions, one must coordinate the spheres such that their exclusion balls overlap as little as geometrically possible while expanding to cover the spatial domain. This is often formalized as the thinnest covering problem \cite{rogers1964packing,Conway1999}.

\textit{Remarks:-} The convergence of both bounds to $\mathcal{P}_I = 1$ for $\phi \rightarrow 0$ is expected as the hard-sphere system approaches an ideal-gas. Analyzing the initial derivatives of the bounds reveals the trajectory towards the fluid state. The initial slope of the upper bound is $
     {d \mathcal{P}^\text{max}_I}/(d \phi)= -{1}/({\phi_{cp}(d)}),$
whereas the initial slope of the lower bound is ${d \mathcal{P}^\text{min}_I}/(d \phi)= -2^d.$ Because $2^d \gg 1/\phi_{cp}(d)$ for all spatial dimensions, the lower bound decreases at a significantly higher rate compared to the upper bound.

In the dilute limit, an equilibrium fluid adheres to this lower bound (Fig. \ref{fig:PI-1d}). When isolated particles are present, the probability of their exclusion zones overlapping approaches zero, meaning the system destroys insertion space at the maximum rate of $2^d$. The equilibrium hard-sphere system initially follows the $1 - 2^d \phi$ trajectory; it does not follow the upper bound, as an ordered crystalline cluster does not occur for a sufficiently dilute gas. Interestingly, in contrast to an equilibrium system, the insertion probability for hard and active rods in $d=1$ appears to follow the upper bound in the limit of infinite activity \cite{Davis2026IS}; thus, the above bounds appears to apply to both active and equilibrium hard spheres.

We now ascertain the form of the surface area of the tight ``exclusion-zone'' cluster ($\omega^\ast$) that provides a close-packed droplet reference. Because internal boundaries are eliminated, the minimum surface area is the outer surface of the tight droplet and is bounded by $S_I \geq dV^\ast / R^\ast$. To evaluate the thermodynamic limit, we define the specific insertion surface area $s_I = S_I / V$. Using \eqref{eq:ClusterRadius} yields $s_I \propto V^{-1/d}$ where in the thermodynamic limit this lower bound converges to zero. Unsurprisingly, the imposed geometry of a tight (spherical) cluster transforms the surface area of the union from an extensive to sub-extensive property.

We can also construct a non-dimensional ratio relating the surface area and volume of the insertion space for the optimized state. Accounting exactly for the finite-size exclusion halo of radius $R^\ast + \sigma$, the surface area of the union is $S_I = d V^\ast (1 + \sigma/R^\ast)^{d-1} / R^\ast$. This yields the exact finite-size ratio:
\begin{equation}
    \frac{S_I \sigma}{V_I} \bigg\rvert_{\omega^\ast} = \frac{\left( d \frac{V^\ast}{R^\ast} \right) \left( 1 + \frac{\sigma}{R^\ast} \right)^{d-1} \sigma}{V - V^\ast \left( 1 + \frac{\sigma}{R^\ast} \right)^d}.
\end{equation}

Dividing the numerator and denominator by $V$, substituting $R^\ast$ \eqref{eq:ClusterRadius}, and expanding for large $N$ isolates the leading-order scaling. Because the boundary layer $\sigma/R^\ast \propto N^{-1/d}$ strictly vanishes in the thermodynamic limit, the finite-size ratio for the spherical-cluster reference asymptotically converges to:
\begin{equation}
    \frac{S_I \sigma}{V_I} \bigg\rvert_{\omega^\ast} \simeq 2 d \left( \frac{\phi_{cp}(d)}{N} \right)^{1/d} \left( \frac{\phi}{\phi_{cp}(d) - \phi} \right),
    \label{eq:S_IOverV_Icluster}
\end{equation}
which is, to the best of our knowledge, a new statistical geometric relation. This scaling confirms that at a fixed density relative to close-packing, $\phi / \phi_{cp}(d)$, the ratio decays as $N^{-1/d}$ in the thermodynamic limit. In the high-dimensional limit ($d \to \infty$), applying the Kabatiansky-Levenshtein bound shows that the term $(\phi_{cp}(d))^{1/d}$ approaches a constant ($\sim 2^{-0.599} \approx 0.6602$). Thus, the prefactor scaling is strictly linear with respect to the spatial dimension, $\mathcal{O}(d)$, reflecting the tendency for high-dimensional spheres to have their volumes concentrated near the surface. We further note that the close-packed spherical cluster underpinning \eqref{eq:S_IOverV_Icluster} implies that $S_I$ and $V_I$ are equivalent to their values per disconnected insertion cavity $s_I$ and $v_I$, respectively.

\section{Hard-sphere thermodynamics}
Connections to thermodynamics begin with the partition function, from which all else is derived. First, we restate an exact relation, originally proved by R.J. Speedy, for the partition function for hard-spheres in terms of products of the insertion volume \cite{Speedy1980}:
\begin{equation}
    Q(N, V, T; d) = \frac{1}{N! \Lambda^{Nd}} \prod_{i=0}^{N-1} \langle V_I(i; d) \rangle,
    \label{eq:speedy_target}
\end{equation}
where $\langle V_I(i; d) \rangle$ is the averaged insertion volume for a system of $i$ spheres in $d$ dimensions. Speedy's relation is remarkably simple and compact, though at some cost of accessible information: all information content of the hard-sphere system, its geometry and any limits such as close-packing, are ``hidden'' in $V_I$.

Now, it will be instructive to provide an upper bound on the partition function \eqref{eq:speedy_target}. From \eqref{eq:PI_upper} the upper bound on the insertion volume for a system of $i$ particles is
\begin{equation}
    V_I(i) \leq V\left(1 - \frac{i v_d}{V \phi_{cp}(d)} \right),
    \label{eq:VI_upper}
\end{equation}
resulting in the finite product:
\begin{equation}
    \begin{aligned}
         \prod_{i=0}^{N-1} \langle V_I(i; d) \rangle \leq  \left( \frac{v_d}{\phi_{cp}(d)}\right)^N \prod_{i=0}^{N-1} \left( \frac{V \phi_{cp}(d)}{v_d} -i\right).
    \end{aligned}
\end{equation}
The remaining product has the form of a falling factorial. Using the fact that $\Gamma(X+1) = X \Gamma(X)$ and the following identity
\begin{equation}
    \begin{aligned}
        \prod_{i=0}^{N-1} (X -i ) &\equiv \frac{\Gamma(X+1)}{\Gamma(X-N+1)},
    \end{aligned}
\end{equation}
allows us to write an exact analytical upper bound on the partition function for the $d$-dimensional hard-sphere system as:
\begin{equation}
\begin{aligned}
    &\qquad \qquad \qquad Q(N,V,T;d) \leq Q^\ast_N, \\
    &Q^\ast_N = \frac{1}{\Lambda^{Nd}} \left( \frac{v_d}{\phi_{cp}(d)} \right)^N \frac{1}{N!} \frac{\Gamma\left( \frac{V \phi_{cp}(d)}{v_d} + 1 \right)}{\Gamma\left( \frac{V \phi_{cp}(d)}{v_d} - N + 1 \right)}.
    \label{eq:raw_Q_bound}
    \end{aligned}
\end{equation}

\begin{figure}[t!]
    \centering
    \includegraphics[width=0.99\linewidth]{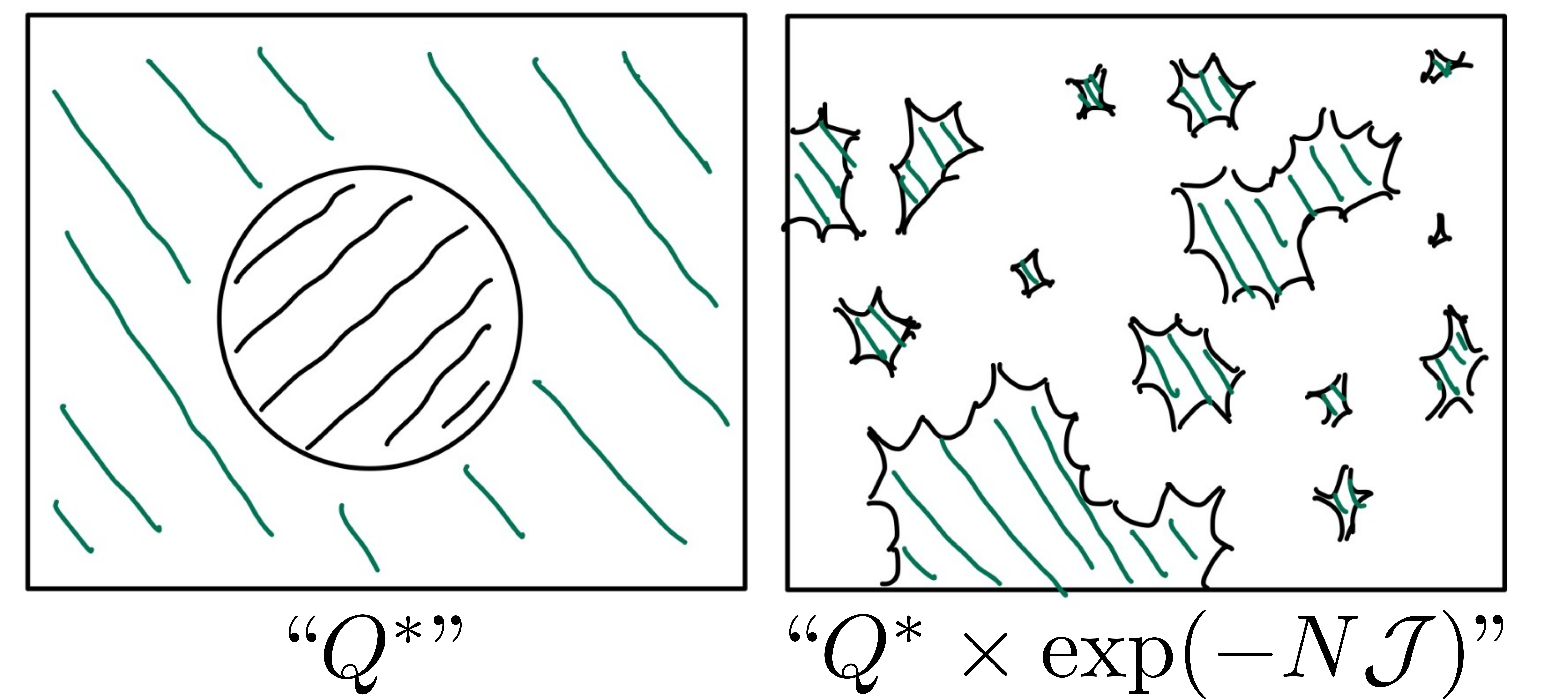}
    \caption{Pictorial representation of the decomposition of the partition function for the hard-sphere system: an upper bound $Q^\ast$ (left), associated with the close-packed droplet reference saturating the insertion-volume bound (green), and a ``discount'' factor $\exp(-N \mathcal{J})$ which gives the precise amount of void-fragmentation for some given packing $\phi$ (right).}
    \label{fig:fig3}
\end{figure}

We now show that \eqref{eq:speedy_target} admits an alternative, but equivalent, exact representation, recovered from \eqref{eq:raw_Q_bound} and written as
\begin{equation}
    Q_N(N,V,T;d) = Q^\ast_N \exp(-N \mathcal{J}_N),
    \label{eq:Q_exactFORM}
\end{equation}
where $\mathcal{J}_N$ is determined below. This representation is, to the best of our knowledge, new, though we stress that it is an exact rearrangement of \eqref{eq:speedy_target}, not an independent solution of the many-body problem.

We, again, start with \eqref{eq:speedy_target} and factor out an upper bound of the insertion volume $V_I^\ast$ as
\begin{equation}
\begin{aligned}
        Q(N, V, T; d) &= \left( \frac{1}{N! \Lambda^{Nd}} \prod_{i=0}^{N-1} V_I^\ast(i) \right) \prod_{i=0}^{N-1} \frac{\langle V_I(i; d) \rangle}{V_I^\ast(i)}, \\
        &= Q^\ast_N \prod_{i=0}^{N-1} \frac{\langle V_I(i; d) \rangle}{V_I^\ast(i)}
    \label{eq:factored_Q}
\end{aligned}
\end{equation}
where $V_I^\ast(i)$ is the right hand side of \eqref{eq:VI_upper}. So, by comparing with \eqref{eq:Q_exactFORM}, we now have the exact finite-$N$ penalty relation:
\begin{equation}
    \exp(-N \mathcal{J}_N) = \prod_{i=0}^{N-1} \frac{\langle V_I(i; d) \rangle}{V_I^\ast(i)}.
    \label{eq:Jdef}
\end{equation}
The intuition behind this factor is that it gives the precise amount of void-fragmentation of the bounding, connected, insertion space (see Fig. \ref{fig:fig3}); this identity holds for any reference $V_I^\ast(i)$, though here we intentionally make the specific choice of \eqref{eq:VI_upper}: the tightest closed-form bound on $\langle V_I(i) \rangle$, realized asymptotically by a compact droplet (see Fig.~\ref{fig:densecluster}).

It will now be useful to take the natural logarithm of the RHS of the above relation, and to re-introduce the insertion probability $\mathcal{P}_I = \langle V_I \rangle/V$, so that we have the exact discrete sum:
\begin{equation}
    \mathcal{J}_N = -\frac{1}{N} \sum_{i=0}^{N-1} \ln \left( \frac{\mathcal{P}_I(\phi_i)}{1 - \phi_i/\phi_{cp}(d)} \right),
    \label{eq:discrete_sum}
\end{equation}
where $\phi_i = i v_d / V$ is the macroscopic packing fraction for $i$ spheres. In the thermodynamic limit ($N \to \infty, V \to \infty$ at fixed $\phi = N v_d / V$), the exact finite-$N$ identity acts as a leading extensive free energy, where the sum converges to a Riemann integral over the packing fraction:
\begin{align}
    \mathcal{J}_N &\simeq -\frac{1}{\phi} \int_0^\phi \ln\left( \frac{\mathcal{P}_I(\phi')}{1 - \phi'/\phi_{cp}(d)} \right) d\phi' \nonumber \\
    &\hspace{-2em}= -\frac{1}{\phi} \int_0^\phi \ln \mathcal{P}_I(\phi') d\phi' + \frac{1}{\phi} \int_0^\phi \ln\left( 1 - \frac{\phi'}{\phi_{cp}(d)} \right) d\phi'.
    \label{eq:integral_split}
\end{align}

We now invoke the identity for the excess (over the ideal gas) chemical potential  $\beta \mu_\text{ex}(N,V,T;d) = - \ln \mathcal{P}_I$\cite{Speedy1980,BOULOUGOURIS1999} so that we may write the following:
\begin{equation}
    \frac{1}{\phi} \int_0^\phi \ln \mathcal{P}_I(\phi')\,d\phi' = -\frac{1}{\phi} \int_0^\phi \beta \mu_\text{ex}(\phi')\,d\phi' = -f_\text{ex}(\phi),
    \label{eq:f_ex_def}
\end{equation}
where $f_\text{ex} = \beta F_\text{ex} / N$.
The excess free energy can be expressed exactly as the integral of the non-ideal compressibility factor $Z$ over the density \cite{Hansen1986}, which reads:
\begin{equation}
    f_\text{ex}(\phi) = \int_0^\phi \frac{Z(\phi') - 1}{\phi'} d\phi'.
    \label{eq:f_ex_integral}
\end{equation}
We now use the following exact identity, again shown by R. J. Speedy, for the compressibility \cite{Speedy1980,Speedy1988}:
\begin{equation}
    Z(\phi') - 1 = \frac{\sigma}{2d} \frac{\langle S_I(\phi') \rangle}{\langle V_I(\phi') \rangle},
    \label{eq:SpeedyEOSRelation}
\end{equation}
which when injected into the expression for $f_\text{ex}(\phi)$ results in
\begin{equation}
    \frac{1}{\phi} \int_0^\phi \ln \mathcal{P}_I(\phi') d\phi' = - \frac{\sigma}{2d} \int_0^\phi \frac{\langle S_I(\phi') \rangle}{\langle V_I(\phi') \rangle} \frac{d\phi'}{\phi'}.
    \label{eq:speedy_substitution}
\end{equation}

Thus, by combining \eqref{eq:speedy_substitution}, \eqref{eq:integral_split}, and \eqref{eq:Jdef} we now arrive at closed-form expression for the thermodynamic limit of $\mathcal{J}$:
\begin{equation}
    \mathcal{J} = \Bigg( \frac{\sigma}{2d} \int_0^\phi \frac{\langle S_I(\phi') \rangle}{\langle V_I(\phi') \rangle} \frac{d\phi'}{\phi'} + \frac{1}{\phi} \int_0^\phi \ln\left( 1 - \frac{\phi'}{\phi_{cp}(d)} \right) d\phi' \Bigg).
    \label{eq:J_integral}
\end{equation}
Substituting this thermodynamic-limit expression into the factorized form and performing the elementary integral in the second term gives the following:
    \begin{align}
     Q(N, V, T; d) &= \frac{1}{\Lambda^{Nd}} \left( \frac{v_d}{\phi_{cp}} \right)^N \frac{1}{N!} \frac{\Gamma\left( \frac{V \phi_{cp}}{v_d} + 1 \right)}{\Gamma\left( \frac{V \phi_{cp}}{v_d} - N + 1 \right)} \times \nonumber \\ 
      &\exp\Bigg[ - N \Bigg( \frac{\sigma}{2d} \int_0^\phi \frac{\langle S_I(\phi') \rangle}{\langle V_I(\phi') \rangle} \frac{d\phi'}{\phi'}  -1 \nonumber \\
      & \qquad \qquad-\frac{\phi_{cp}-\phi}{\phi} \ln\left(1 - \frac{\phi}{\phi_{cp}} \right) \Bigg) \Bigg].
     \label{eq:Q_explicit_exact}
\end{align}
One is free to exchange $\langle S_I(\phi') \rangle/\langle V_I(\phi') \rangle$ with $\langle s_f (\phi')/v_f (\phi')\rangle$: the ensemble (cavity) averaged ratio of the boundary of the local free-volume cavity to its volume \cite{Speedy1980,Sastry1998}. With such a substitution, the thermodynamic-limit representation can be written solely in terms of local (microscopic) free-volume information. In Appendix B (SI) we verify that \eqref{eq:Q_explicit_exact} recovers the Tonks' solution ($d=1$), and in Appendix C we show the internal consistency with the equation of state \eqref{eq:SpeedyEOSRelation}.

For the low- and high-density regimes of \eqref{eq:Q_explicit_exact} we find the following approximate partition functions:
\begin{align}
       Q(N, V, T; d) &\underset{\phi \ll \phi_{cp} }{\simeq} \frac{V^N}{\Lambda^{Nd} N!} (1 - 2^d \phi)^{N/2},
\end{align}
for the low density regime, and
\begin{align}
 Q(N, V, T; d) &\asymp \frac{1}{\Lambda^{Nd}} \left( \frac{v_d}{\phi_{cp}} \right)^N C^N \left( 1 - \frac{\phi}{\phi_{cp}} \right)^{Nd},
\end{align}
for the high density regime (see Appendix D in SI). 

For clarity: (i) the finite-$N$ identities \eqref{eq:speedy_target}, \eqref{eq:factored_Q}--\eqref{eq:discrete_sum} are exact for any $N$; (ii) Eqs.~\eqref{eq:J_integral}--\eqref{eq:Q_explicit_exact} hold in the thermodynamic limit; (iii) the sharp bound \eqref{eq:PI_upper} and $Q_N^\ast$ are conditional on a certificate at $\Delta_d=\phi_{cp}(d)$, established only for $d=1,8,24$; and (iv) the low/high-density forms are asymptotic approximations.

At low densities ($\phi \ll \phi_{cp}$), the partition function reduces to the ideal continuum gas while generating the exact leading-order corrections \cite{Hansen1986}. These include, the binomial expansion $(1 - 2^d \phi)^{N/2} \simeq \exp(-N 2^{d-1} \phi)$ yields the classical second virial coefficient, $B_2 = 2^{d-1}$. Absorbing this term transforms the volume into $(V - N b)^N$, where $b = 2^{d-1}v_d$ is the van der Waals excluded volume \cite{Reif1965}. At high densities ($\phi \to \phi_{cp}$), the partition function reduces to the classical free-volume (cell) theory of the solid state \cite{Hoover1966}. With the close-packed cell volume $v_{cp} = v_d/\phi_{cp}$ and specific volume $v = V/N$, the density term scales as $(1 - \phi/\phi_{cp})^{Nd} \propto (v - v_{cp})^{Nd}$, recovering the known simple pole divergence \cite{Salsburg1966}. Furthermore, as the close-packed ``capacity'' goes like $N$, the bound $Q^\ast$ absorbs the $1/N!$ indistinguishability factor, leaving an exponential prefactor $C^N$ dependent on the nonuniversal finite parts of the free-volume integral. This mathematically reflects the fracturing of the--otherwise continuous--void space, where particles are confined to \textit{distinguishable} cages.

\section{Discussion}
We have here presented an alternative exact finite-$N$ factorization of the hard-sphere partition function, together with its thermodynamic-limit form. This formulation involves factoring out the geometric upper bound of the available insertion space (the close-packed state), $Q^\ast$, as opposed to the usual approach of factoring out the ideal gas limit. This approach provides a distinct theoretical viewpoint. The equilibrium fluid is treated mathematically as a topologically fragmented solid rather than a condensed gas. The resulting form of the partition function shows that understanding the thermodynamics of the hard-sphere system is, then, essentially about working out how the continuous/connected void space fragments as a function of density, a process captured precisely by the geometric penalty $\mathcal{J}$.

Although $\mathcal{J}(\phi)$ carries the same many-body content as $\mathcal{P}_I(\phi)$ and is not itself a solution of the hard-sphere problem, expressing it through $\langle S_I \rangle/\langle V_I \rangle$ is advantageous: by the exact identity \eqref{eq:SpeedyEOSRelation} this ratio is proportional to $Z(\phi)-1$, so $\mathcal{J}(\phi)$ can be evaluated from any known equation of state with no further geometric computation. Computationally, determining the geometric integral in $\mathcal{J}$ might initially appear more unwieldy than evaluating Speedy's exact discrete product over insertion volumes. However, we argue \textit{a priori} that it offers a  statistical advantage. Factoring the geometric bounds out of $\langle V_I(i) \rangle$ removes the computational burden of adequately sampling exponentially rare, macroscopic insertion cavities at high densities. Furthermore, the resulting integrand, $\langle S_I(\phi') \rangle / \langle V_I (\phi')\rangle \equiv \langle s_f/v_f \rangle(\phi')$, has been shown to be a more statistically controlled/robust quantity \cite{Sastry1997}. Because fluctuations that increase the volume of a void inherently increase its bounding surface area, this covariance suppresses the runaway variance arising from direct sampling of the insertion volume alone \cite{Speedy1991}.

Overall, we have made some progress in the statistical geometric understanding of the hard-sphere system. Our work is a clear reminder that all thermodynamic content of the hard-sphere system rests in determining the averaged surface to volume ratio of the union of the excluded-volume \cite{Speedy1980,Reiss1992}, and motivates further mathematical and theoretical work to determine it. Other future work may include the direct numerical evaluation of \eqref{eq:Q_explicit_exact} via computational statistical geometry, including $\mathcal{J}(\phi)$ in $d=2,3$ from existing free-volume data \cite{Sastry1997,Sastry1997ii,Sastry1998,Davis2026IS,Speedy1991}, further exploring the consequences of the geometric and thermodynamic bounds for active hard spheres \cite{ArnoulxdePirey2019,Davis2026IS}, and connecting statistical geometry concepts, and relations, with those of the replica theory \cite{Parisi2006,Parisi2010}.

\acknowledgments
The author thanks Ian J. Ford for discussions. L.K.D. is funded by a Flora Philip Fellowship at the University of Edinburgh.

\bibliographystyle{eplbib} 
\bibliography{refs}



\section*{Supplementary material}

sharp bound involving $\phi_{cp}(d)$ follows whenever such a certificate is available at the optimal density. More generally, if a Blichfeldt-type local density certificate at density scale $\Delta_d$ is available, the same argument gives the weaker but rigorous bound $\mathcal{P}_I \le 1 - \phi/\Delta_d$.

Let $X_N = \{\mathbf{r}_1, \ldots, \mathbf{r}_N\}$ be a hard-sphere configuration with

$$|\mathbf{r}_i - \mathbf{r}_j| \ge \sigma, \tag{A1}$$

where $\sigma$ is the physical hard-sphere diameter. The physical single-particle volume is $v_d$. A trial particle center is excluded from the ball

$$\mathcal{B}_\sigma(\mathbf{r}_i) = \{\mathbf{r} : |\mathbf{r} - \mathbf{r}_i| < \sigma\}, \tag{A2}$$

which has radius $\sigma$ and diameter $2\sigma$. The total exclusion set is therefore

$$E(X_N) = \bigcup_{i=1}^{N} \mathcal{B}_\sigma(\mathbf{r}_i). \tag{A3}$$

The insertion volume and insertion probability are

$$V_I(X_N) = V - \mathrm{Vol}(E(X_N)), \tag{A4}$$

and

$$\mathcal{P}_I(X_N) = 1 - \frac{\mathrm{Vol}(E(X_N))}{V}. \tag{A5}$$

Thus, to prove an upper bound on $\mathcal{P}_I$, it is sufficient to prove a lower bound on $\mathrm{Vol}(E(X_N))$.

Assume there exists a non-negative local density certificate $g : \mathbb{R}^d \to [0, \infty)$ satisfying

1. $\mathrm{supp}\, g \subseteq \mathcal{B}_\sigma(\mathbf{0})$,
2. the following integral
$$\int_{\mathbb{R}^d} g(\mathbf{r})\, d\mathbf{r} = \frac{v_d}{\Delta_d}, \tag{A6}$$
where $\Delta_d$ is a density scale for which the present local certificate exists,
3. and that for every hard-sphere packing $X$,
$$\sum_{\mathbf{r}_i \in X} g(\mathbf{r} - \mathbf{r}_i) \le 1 \qquad \text{for all } \mathbf{r} \in \mathbb{R}^d. \tag{A7}$$

This is a Blichfeldt-type certificate: it assigns to every particle a smeared cell volume $v_d/\Delta_d$, while ensuring that the total smeared density never exceeds unity.

For the finite configuration $X_N$, define

$$\rho_g(\mathbf{r}) = \sum_{i=1}^{N} g(\mathbf{r} - \mathbf{r}_i). \tag{A8}$$

Because $g$ is supported inside the exclusion ball, $\rho_g(\mathbf{r}) = 0$ outside $E(X_N)$. Therefore

$$\int_{\mathbb{R}^d} \rho_g(\mathbf{r})\, d\mathbf{r} = \int_{E(X_N)} \rho_g(\mathbf{r})\, d\mathbf{r}. \tag{A9}$$

By the pointwise certificate condition, $\rho_g(\mathbf{r}) \le 1$, and hence

$$\int_{E(X_N)} \rho_g(\mathbf{r})\, d\mathbf{r} \le \mathrm{Vol}(E(X_N)). \tag{A10}$$

On the other hand,

$$\int_{\mathbb{R}^d} \rho_g(\mathbf{r})\, d\mathbf{r} = \sum_{i=1}^{N} \int_{\mathbb{R}^d} g(\mathbf{r} - \mathbf{r}_i)\, d\mathbf{r} = \frac{N v_d}{\Delta_d}. \tag{A11}$$

Combining the above identities gives

$$\mathrm{Vol}(E(X_N)) \ge \frac{N v_d}{\Delta_d}. \tag{A12}$$

Consequently,

$$\mathcal{P}_I(X_N) \le 1 - \frac{N v_d}{V \Delta_d} = 1 - \frac{\phi}{\Delta_d}. \tag{A13}$$

If the certificate is sharp, so that $\Delta_d = \phi_{cp}(d)$, this becomes

$$\mathcal{P}_I(X_N) \le 1 - \frac{\phi}{\phi_{cp}(d)}. \tag{A14}$$

For a periodic system, the same proof applies to the periodically lifted packing, with the periodic density field integrated over a single fundamental cell.

Finally, the bound is asymptotically sharp. Consider a sequence of saturated close-packed clusters $X_R$ obtained by restricting a packing of density $\phi_{cp}(d)$ to a window $W_R$ of linear size $R$. Then

$$N_R v_d = \phi_{cp}(d)\, \mathrm{Vol}(W_R) + \mathcal{O}(R^{d-1}). \tag{A15}$$

Since the packing is saturated, the exclusion balls cover the bulk of $W_R$, with only boundary corrections. Therefore

$$\begin{aligned} \mathrm{Vol}(E(X_R)) &= \mathrm{Vol}(W_R) + \mathcal{O}(R^{d-1}), \\ &= \frac{N_R v_d}{\phi_{cp}(d)} + \mathcal{O}(N_R^{(d-1)/d}). \end{aligned} \tag{A16}$$

Placing this cluster in a total volume $V = N_R v_d/\phi$ gives

$$\mathcal{P}_I(X_R) = 1 - \frac{\phi}{\phi_{cp}(d)} + \mathcal{O}(N_R^{-1/d}). \tag{A17}$$

Hence

$$\lim_{N\to\infty} \sup_{X_N} \mathcal{P}_I(X_N) = 1 - \frac{\phi}{\phi_{cp}(d)}, \tag{A18}$$

provided the sharp upper certificate with $\Delta_d = \phi_{cp}(d)$ is available.

### Appendix B: Recovering the exact partition function in $d = 1$

We now show how, in $d = 1$, (36) reduces to the exactly known partition function for a fluid of hard rods of length $\sigma$. With $v_1 = \sigma$ and $\phi_{cp} = 1$, the ratio of Gamma functions defining the close-packed bound, $Q^*$, expands exactly into the following:

$$\frac{\Gamma(V/\sigma+1)}{\Gamma(V/\sigma-N+1)} = \prod_{i=0}^{N-1}\left(\frac{V}{\sigma}-i\right) = \sigma^{-N}\prod_{i=0}^{N-1}(V-i\sigma). \tag{B1}$$

The prefactor, in (25), is therefore $Q^* = (\Lambda^N N!)^{-1}\prod_{i=0}^{N-1}(V - i\sigma)$. In the thermodynamic limit, we are permitted to write:

$$\begin{aligned}\ln Q^* &\simeq \ln\left(\frac{V^N}{\Lambda^N N!}\right) + N\int_0^1 \ln(1-x\phi)dx \\ &= \ln\left(\frac{V^N}{\Lambda^N N!}\right) - N\left(1+\frac{1-\phi}{\phi}\ln(1-\phi)\right).\end{aligned} \tag{B2}$$

We immediately recognise the first term as the logarithm of the partition function of a one-dimensional ideal gas.

Next, we evaluate the logarithm of the second factor $\ln Q_{\mathcal{J}} = -N\mathcal{J}(\phi)$. For a 1D continuum, the insertion boundary and volume are governed strictly by the macroscopic probability of an interparticle contact bond, $P_C$. In one-dimension the insertion boundary is simply twice the number of insertion cavities $\langle S_I\rangle = 2\langle N_I\rangle = 2N(1-P_C)$. The expected insertion volume is exactly given as $\langle V_I\rangle = (V - N\sigma)(1-P_C)$. Taking the ratio $\langle S_I\rangle/\langle V_I\rangle$ results in the cancellation of factors of $1 - P_C$ and leads to the following:

$$\frac{\sigma}{2}\frac{\langle S_I(\phi)\rangle}{\langle V_I(\phi)\rangle} = \frac{\sigma N}{V-N\sigma} = \frac{\phi}{1-\phi}. \tag{B3}$$

Substituting this ratio into $\mathcal{J}$ gives:

$$\mathcal{J}(\phi) = \int_0^\phi \frac{1}{\phi'}\left(\frac{\phi'}{1-\phi'}\right)d\phi' + \frac{1}{\phi}\int_0^\phi \ln(1-\phi')d\phi', \tag{B4}$$

which can be evaluated as:

$$\begin{aligned}\mathcal{J}(\phi) &= -\ln(1-\phi) - \left[1+\frac{1-\phi}{\phi}\ln(1-\phi)\right], \\ &= -1 - \frac{\ln(1-\phi)}{\phi}.\end{aligned} \tag{B5}$$

The exact logarithm of the partition function is the sum $\ln Q = \ln Q^* - N\mathcal{J}(\phi)$, and so summing (B2) and (B5) leads to:

$$\begin{aligned}\ln Q &= \ln\left(\frac{V^N}{\Lambda^N N!}\right) - N\left(1+\frac{1-\phi}{\phi}\ln(1-\phi)\right) \\ &\quad - N\left(-1-\frac{\ln(1-\phi)}{\phi}\right) \\ &= \ln\left(\frac{V^N}{\Lambda^N N!}\right) + N\ln(1-\phi).\end{aligned} \tag{B6}$$

Absorbing the exponential argument resolves the total exact integral:

$$\ln Q = \ln\left(\frac{V^N(1-\phi)^N}{\Lambda^N N!}\right). \tag{B7}$$

Substituting the definition of the packing fraction, $\phi = N\sigma/V$, we recover the exact partition function for the Tonks fluid [3]:

$$Q(N,V,T;1) = \frac{(V-N\sigma)^N}{\Lambda^N N!}. \tag{B8}$$

### Appendix C: Internal consistency with the equation of state

To verify the thermodynamic consistency of the exact partition function, we recover the continuous statistical-geometric equation of state [19]. The compressibility factor $Z = PV/(Nk_BT)$ is defined via the volume derivative of the partition function:

$$Z = \frac{V}{N}\left(\frac{\partial \ln Q}{\partial V}\right)_{N,T}. \tag{C1}$$

Transforming the derivative with respect to the packing fraction $\phi = Nv_d/V$ yields $(\partial\phi/\partial V)_N = -\phi/V$, such that:

$$Z(\phi) = -\frac{\phi}{N}\left(\frac{\partial \ln Q}{\partial \phi}\right)_{N,T}. \tag{C2}$$

We decompose the logarithm of the partition function into the close-packed bound, $\ln Q^*$, and the continuous geometric penalty, $\ln Q_{\mathcal{J}} = -N\mathcal{J}(\phi)$. The close-packed bound is:

$$\ln Q^* = \ln\mathcal{C}^* + \ln\Gamma(M+1) - \ln\Gamma(M-N+1), \tag{C3}$$

where $\mathcal{C}^*$ is a volume-independent constant and $M = V\phi_{cp}/v_d = N\phi_{cp}/\phi$ is the total capacity of close-packed sites. Applying Stirling's approximation, $\ln\Gamma(x+1) \simeq x\ln x - x$, in the thermodynamic limit gives:

$$\ln Q^* \simeq \ln\mathcal{C}^* + M\ln M - (M-N)\ln(M-N) - N. \tag{C4}$$

Taking the partial derivative with respect to $M$ yields:

$$\frac{\partial \ln Q^*}{\partial M} = -\ln\left(1-\frac{N}{M}\right). \tag{C5}$$

Substituting $N/M = \phi/\phi_{cp}$ and applying the chain rule, $(\partial M/\partial V) = \phi_{cp}/v_d$, isolates the close-packed contribution to the compressibility:

$$Z^* = \frac{V}{N}\frac{\phi_{cp}}{v_d}\left[-\ln\left(1-\frac{\phi}{\phi_{cp}}\right)\right] = -\frac{\phi_{cp}}{\phi}\ln\left(1-\frac{\phi}{\phi_{cp}}\right). \tag{C6}$$

The penalty contribution to the compressibility is $Z_{\text{penalty}} = \phi(\partial\mathcal{J}/\partial\phi)$. The integral expression for $\mathcal{J}(\phi)$ is:

$$\mathcal{J}(\phi) = \frac{\sigma}{2d}\int_0^{\phi} \frac{\langle S_I(\phi')\rangle}{\langle V_I(\phi')\rangle}\frac{d\phi'}{\phi'} + \frac{1}{\phi}\int_0^{\phi} \ln\left(1-\frac{\phi'}{\phi_{cp}}\right)d\phi'. \tag{C7}$$

Evaluating the density derivative using standard calculus and multiplying by $\phi$ yields:

$$Z_{\text{penalty}} = \frac{\sigma}{2d}\frac{\langle S_I(\phi)\rangle}{\langle V_I(\phi)\rangle} - \frac{1}{\phi}\int_0^{\phi}\ln\left(1-\frac{\phi'}{\phi_{cp}}\right)d\phi' + \ln\left(1-\frac{\phi}{\phi_{cp}}\right). \tag{C8}$$

Substituting the exact algebraic identity for the remaining integral, $\frac{1}{\phi}\int_0^{\phi}\ln(1-\frac{\phi'}{\phi_{cp}})d\phi' = -1-\frac{\phi_{cp}-\phi}{\phi}\ln(1-\frac{\phi}{\phi_{cp}})$, simplifies the penalty to:

$$Z_{\text{penalty}} = \frac{\sigma}{2d}\frac{\langle S_I(\phi)\rangle}{\langle V_I(\phi)\rangle} + 1 + \frac{\phi_{cp}}{\phi}\ln\left(1-\frac{\phi}{\phi_{cp}}\right). \tag{C9}$$

The total compressibility factor is the sum $Z(\phi) = Z^* + Z_{\text{penalty}}$. Summing (C6) and (C9) exactly cancels the logarithmic terms:

$$Z(\phi) = 1 + \frac{\sigma}{2d}\frac{\langle S_I(\phi)\rangle}{\langle V_I(\phi)\rangle}. \tag{C10}$$

confirming the thermodynamic consistency of (36).

## Appendix D: Low and high density regimes

We now probe the exact partition function (36) in the low and high density regimes. First, we probe the low density regime ($\phi \ll \phi_{cp}$). We first evaluate the discrete combinatorial prefactor. Let $M = V\phi_{cp}/v_d$ represent the continuous capacity of close-packed sites. The ratio of the Gamma functions expands asymptotically for $N \ll M$:

$$\begin{aligned}\ln\left[\frac{\Gamma(M+1)}{\Gamma(M-N+1)}\right] &= \sum_{i=0}^{N-1}\ln(M-i) \\ &\simeq N\ln M - \sum_{i=0}^{N-1}\frac{i}{M} \\ &\simeq N\ln M - \frac{N^2}{2M}.\end{aligned} \tag{D1}$$

Substituting the physical definition $N/M = \phi/\phi_{cp}$ back into the scalar prefactors of the bounding partition function, $Q^*$, isolates its low-density scaling:

$$Q^* \simeq \frac{V^N}{\Lambda^{Nd}N!}\exp\left(-\frac{N\phi}{2\phi_{cp}}\right). \tag{D2}$$

Taylor expanding the second term in $\mathcal{J}$ (35) to first order in $\phi/\phi_{cp}$ yields:

$$\begin{aligned}\frac{1}{\phi}\int_0^{\phi}\ln\left(1-\frac{\phi'}{\phi_{cp}}\right)d\phi' &= -1 - \frac{\phi_{cp}-\phi}{\phi}\ln\left(1-\frac{\phi}{\phi_{cp}}\right) \\ &\simeq -1 - \left(\frac{\phi_{cp}}{\phi}-1\right)\left(-\frac{\phi}{\phi_{cp}} - \frac{\phi^2}{2\phi_{cp}^2}\right) \\ &\simeq -\frac{\phi}{2\phi_{cp}}.\end{aligned} \tag{D3}$$

In the partition function this term is scaled by a factor of $-N$. Multiplying this continuous penalty by the discrete bound in Eq. (D2) results in the following cancellation of the $\phi_{cp}$ terms: $\exp(-N\phi/2\phi_{cp})\exp(+N\phi/2\phi_{cp}) = 1$. This guarantees that the reference state of the fluid is strictly an ideal (continuous) gas, with no hint of any close-packed structure.

Remaining imperfections on the ideal gas will now come from the first term in $\mathcal{J}$. For a sufficiently dilute system of spheres of diameter $\sigma$, the insertion cavity is constrained by isolated exclusion spheres of diameter $2\sigma$. The expected available volume is the total space minus the independent exclusion zones, $\langle V_I\rangle \simeq V - Nv_d(2\sigma)$. The bounding surface area is the sum of these independent exclusion envelopes, $\langle S_I\rangle \simeq Na_d(2\sigma)$.

From basic $d$-dimensional geometry, the surface area $a_d(r)$ of a $d$-sphere is related to its volume $v_d(r)$ by $a_d(r) = (d/r)v_d(r)$. For the exclusion radius $r=\sigma$, the surface area is exactly $a_d(2\sigma) = (d/\sigma)v_d(2\sigma)$. Then, the insertion surface-to-volume ratio simplifies precisely to:

$$\begin{aligned}\frac{\sigma}{2d}\frac{\langle S_I\rangle}{\langle V_I\rangle} &\simeq \frac{\sigma}{2d}\left[\frac{N\frac{d}{\sigma}v_d(2\sigma)}{V-Nv_d(2\sigma)}\right] \\ &= \frac{1}{2}\left[\frac{Nv_d(2\sigma)}{V-Nv_d(2\sigma)}\right].\end{aligned} \tag{D4}$$

Next, we rewrite this solely in terms of the packing fraction, $\phi = Nv_d(\sigma)/V$. Because the volume of the exclusion sphere scales as $v_d(2\sigma) = 2^d v_d(\sigma)$, the exclusion volume fraction is $Nv_d(2\sigma)/V = 2^d\phi$ resulting in:

$$\frac{\sigma}{2d}\frac{\langle S_I(\phi)\rangle}{\langle V_I(\phi)\rangle} \simeq \frac{2^{d-1}\phi}{1-2^d\phi}. \tag{D5}$$

Substituting this into the integral in $\mathcal{J}$ and evaluating gives:

$$\begin{aligned}\int_0^{\phi}\left(\frac{2^{d-1}\phi'}{1-2^d\phi'}\right)\frac{d\phi'}{\phi'} &= \int_0^{\phi}\frac{2^{d-1}}{1-2^d\phi'}d\phi' \\ &= -\frac{1}{2}\ln(1-2^d\phi).\end{aligned} \tag{D6}$$

Applying this final exponential geometric penalty to the ideal continuum gas yields the low-density partition function:

$$Q(N,V,T;d) \simeq \frac{V^N}{\Lambda^{Nd}N!}(1-2^d\phi)^{N/2}. \tag{D7}$$

Expanding the binomial limit $(1-x)^{N/2} \simeq \exp(-Nx/2)$ and substituting $2^d\phi = Nv_d(2\sigma)/V$, the exponential correction gives back a restricted insertion volume:

$$V^N \exp\left(-\frac{N^2 v_d(2\sigma)}{2V}\right) \simeq \left(V - \frac{N}{2} v_d(2\sigma)\right)^N = (V - Nb)^N, \quad \text{(D8)}$$

where $b = 2^{d-1}v_d(\sigma)$ is the classical van der Waals excluded volume [46], arrived at without any appeal to the virial.

Note that, (D5) suggests a low-density, $d$-dimensional, equation of state of the form

$$Z(\phi) = 1 + \frac{2^{d-1}\phi}{1-2^d\phi} \equiv \frac{1-2^{d-1}\phi}{1-2^d\phi} = \frac{1-2^{d-1}\phi}{\mathcal{P}'_I(\phi,d)}, \quad \text{(D9)}$$

with a denominator that is precisely the non-zero lower bound of the insertion probability $\mathcal{P}'_I(\phi,d) = 1-2^d\phi$ (15). At fixed scaled density $\tilde{\phi} = 2^d\phi$, this Padé-like form gives $Z \simeq 1 + [\tilde{\phi}/2]/(1-\tilde{\phi})$, while the ideal-gas limit is recovered only when $\tilde{\phi} \ll 1$.

Next, we probe the high-density regime ($\phi \to \phi_{cp}$). As the fluid compresses into a solid state, we expect the available volume becomes restricted to individual particle cages. Let $\epsilon = 1 - \phi/\phi_{cp}$. The exact integral in $\mathcal{J}$ evaluates to:

$$\frac{1}{\phi}\int_0^\phi \ln\left(1-\frac{\phi'}{\phi_{cp}}\right)d\phi' = -1 - \frac{\phi_{cp}-\phi}{\phi}\ln\left(1-\frac{\phi}{\phi_{cp}}\right) = -1 - \frac{\phi_{cp}}{\phi}\epsilon\ln\epsilon \underset{\epsilon\to 0}{=} -1. \quad \text{(D10)}$$

Conversely, the boundary integral strictly diverges. In the asymptotic solid limit, theory dictates that the equation of state diverges with a simple pole proportional to the spatial dimensionality, $Z(\phi) \simeq d/(1-\phi/\phi_{cp})$ [49]. Substituting this into the exact geometric identity isolates the scaling of the internal boundaries:

$$\frac{\sigma}{2d}\frac{\langle S_I(\phi)\rangle}{\langle V_I(\phi)\rangle} \simeq \frac{d}{1-\frac{\phi}{\phi_{cp}}}. \quad \text{(D11)}$$

Integrating this spatial ratio over the density path isolates the dominant logarithmic divergence at the upper limit:

$$\int^\phi \frac{d}{1-\frac{\phi'}{\phi_{cp}}}\frac{d\phi'}{\phi'} \simeq -d\ln\left(1-\frac{\phi}{\phi_{cp}}\right). \quad \text{(D12)}$$

The singular part of the boundary integral gives

$$\mathcal{J}_{\text{sing}}(\phi) \simeq -d\ln\left(1-\frac{\phi}{\phi_{cp}}\right). \quad \text{(D13)}$$

Regular finite terms in $\mathcal{J}$ contribute only an exponential constant $C^N$. Therefore

$$\exp[-N\mathcal{J}(\phi)] \asymp C^N\left(1-\frac{\phi}{\phi_{cp}}\right)^{Nd}. \quad \text{(D14)}$$

Finally, we evaluate the discrete close-packed bound $Q^*$ as the system volume $V$ approaches the terminal close-packed volume $V_{cp} = Nv_d/\phi_{cp}$. The capacity then goes like $M \to N$, and the entropy vanishes:

$$\ln\left[\frac{\Gamma(M+1)}{N!\Gamma(M-N+1)}\right] \simeq M\ln\left(\frac{M}{N}\right) - (M-N)\ln\left(\frac{M-N}{N}\right) \to 0. \quad \text{(D15)}$$

Because there is only one way to assign $N$ identical particles to $N$ localized close-packed sites, the $1/N!$ indistinguishability factor goes away. Thus, the prefactor $Q^*$ simplifies as:

$$Q^* \simeq \frac{1}{\Lambda^{Nd}}\left(\frac{v_d}{\phi_{cp}}\right)^N. \quad \text{(D16)}$$

The dominant singular scaling is thus $Q \propto (1-\phi/\phi_{cp})^{Nd}$. The exponential prefactor is sensitive to nonuniversal finite parts of the free-volume integral and should not be inferred from the leading pole alone. We therefore express the high-density partition function as a scaling form:

$$Q(N,V,T;d) \asymp \frac{1}{\Lambda^{Nd}}\left(\frac{v_d}{\phi_{cp}}\right)^N C^N\left(1-\frac{\phi}{\phi_{cp}}\right)^{Nd}, \quad \text{(D17)}$$

where the constant $C$ absorbs the regular parts of the integration.



\end{document}